# Study of Magnetic properties of $A_2B$ $NbO_6$ (A=Ba, Sr, (BaSr); and B = Fe and Mn) double perovskites

**N.Rama[a,b], J.B.Philipp[c], M.Opel[c], K.Chandrasekaran[a], V.Sankaranarayanan[b], R.Gross[c] and M.S.Ramachandra Rao[a,b#]**

a Materials Science Research Centre, Indian Institute of Technology Madras, Chennai-600036, India

b Department of Physics, Indian Institute of Technology Madras, Chennai-600036, India

c Walther-Meissner Institute, Walther-Meissner Str 8, Garching, D-85748, Germany

## Abstract

We have studied the magnetic properties of $Ba_2FeNbO_6$ and $Ba_2MnNbO_6$. It is seen that $Ba_2FeNbO_6$ is an antiferromagnet with a weak ferromagnetic behaviour at 5K while $Ba_2MnNbO_6$ shows two magnetic transitions one at 45K and the other at 12K. Electron spin resonance (ESR) measurements at room temperature show that the Mn compound does not show any Jahn-Teller distortion. It is also seen that the Neel temperatures of the $A_2FeNbO_6$ (A=Ba, Sr, BaSr) compounds do not vary significantly. However variations in the average A-site ionic radius influence the formation of short range correlations that persists above $T_N$.

---

#Corresponding Author: msrrao@iitm.ac.in , msrrao@squid.umd.edu , Present address: Materials Research Science and Engineering Centre (MRSEC) and Department of Physics, University of Maryland, College Park MD 20742, USA

# INTRODUCTION

Double perovskites of the general form $A_2B'B''O_6$ (A=Sr, Ca. Ba; B'=Fe, Mn, Cr; B''= Mo, W, Ta, Nb) are actively being investigated in order to understand the nature of magnetism and magnetotransport in these compounds. It is known that the B site can accommodate two kinds of metal ions. The cation arrangement on the B-sublattice of a double perovskite is controlled primarily by the charge difference between the B cations and secondarily by the ionic size difference between them. Depending on the valencies and the ionic radii, the B cations settle either orderly or in a random fashion in the lattice [1]. Therefore, the B cations generally determine the physical properties of perovskites. Different kinds of B and B ion pairs show a variety of physical properties of ordered perovskites. For example $Sr_2FeMoO_6$ is a ferromagnetic (FM) metal with a $T_C$ of about 400K [2,3], while $Sr_2MnMoO_6$ is an antiferromagnetic (AFM) insulator with a $T_N$ of 12K [3]. With this in mind we have studied the magnetic properties of $Ba_2B'NbO_6$ (B'= Fe, Mn) (BFNO, BMNO respectively) to determine differences in their magnetic properties, if any, arising because of $Mn^{3+}$ (Jahn Teller ion) and $Fe^{3+}$ ion. We have also studied the magnetic behaviour of $A_2FeNbO_6$ with A=Ba, Sr and BaSr to check the influence of A-site ion on the magnetic properties.

# EXPERIMENT

The samples were prepared using conventional solid-state method. High pure (99.9 %) carbonates of the alkaline earths $BaCO_3$ and $SrCO_3$, oxides of the transition metals $Fe_2O_3$, $MnO_2$ and $Nb_2O_5$ were taken and weighed stoichiometrically. The weighed powders were intimately mixed using an agate mortar and calcined at 1200°C for 24h.This procedure was repeated three times and the resultant powder was pressed into pellets and sintered at 1350°C. The phase purity was checked using x-ray diffraction (Rich-Seifert, Germany). The temperature variation of DC magnetization was done using a SQUID magnetometer (MPMS XL, Quantum Design) in the field-cooled mode (FC). Electron spin resonance measurements were done at room temperature using a X-band Varian Spectrometer.

## RESULTS AND DISCUSSIONS

The X-ray diffractograms of BFNO and BMNO are given in figure1. The samples could be indexed to a cubic lattice with lattice constants of 8.11Å and 8.12Å respectively. Among, the A-site substituted compounds the Sr substituted compound SFNO shows orthorhombic structure (a=5.62Å, b=7.96Å c=5.61 Å) while the BaSr (BSFNO) compound is cubic (8.031Å).

The temperature variation of magnetization for BFNO is shown in figure 2. The temperature derivative of magnetization shows a minimum at 25K (Inset in figure 2). Though, $1/\chi$ vs T is linear (figure3) above 200K, the value of the effective Bohr magneton number obtained was 35.8 $\mu_B$ as against the expected spin only moment of $Fe^{3+}$ (5.92 $\mu_B$ -high spin configuration or 3.37 $\mu_B$ - low spin configuration). Magnetic hysteresis is seen at 5K (figure 4) revealing the presence of weak FM behaviour.

The temperature derivative of magnetization for the Mn-substituted compound (Right inset in figure 2) shows two distinct minima, one at 45K and the other at 12K. The minimum at low temperature is attributed to the Neel temperature (Left inset in figure 2). To determine the origin of high temperature minimum we referred to the neutron and magnetization measurements on $Ba_2MnWO_6$ by Azad et.al. [4], which shows a Neel temperature at 9K and a canted antiferromagnetic like state below 45K. Thus comparing Azad et al's [4] data to ours we attribute the high temperature minimum corresponding to the shoulder in the M-T plot to a canted antiferromagnetic state. Curie-Weiss behaviour is seen at high temperatures (above 200K), which yields a Curie constant of 3.33 emu-mol$^{-1}$K$^{-1}$ and a Weiss temperature of -59K. The negative sign indicates antiferromagnetic behaviour. The value of effective Bohr magneton number calculated from the fit yields a value of 5.18 $\mu_B$, which is comparable to the expected spin only value for $Mn^{3+}$ equal to 4.9 $\mu_B$ in contrast to the effective Bohr magneton number obtained for BFNO pointing to the latter's complex paramagnetic state. There is no magnetic hysteresis seen at 5K indicating that there is a complete AFM ordering below $T_N$ unlike BFNO.

We have performed electron paramagnetic resonance studies on both the Mn and Fe compounds at room temperature. A single broad Lorentzian line is observed with the centre field coinciding with that of the 'g' marker diphenyl-ß-picrylhydrazyl (DPPH) (figure 5). Hence the value of 'g' is nearly 2.00, which is the value for paramagnetic $Fe^{3+}$ ions in CaO [5]. The observed value of 2.00 for the Mn sample is also consistent with the reported value of 'g' obtained for $Mn^{3+}$ ions in $TiO_2$ [6]. This suggests that 'g' value is nearly isotropic and the $MnO_6$ octahedra are undistorted. Comparing this with $LaMnO_{3.00}$, which exhibits a co-operative Jahn-Teller distortion below 600K where a 'g' value of 1.98 is obtained in the paramagnetic state ($T_N$ =140K) due to the distortion of $MnO_6$ octahedra [7], we see that in $Ba_2MnNbO_6$ such a distortion is absent. This is also supported from our X-ray diffraction data where we see only a cubic symmetry. In contrast, compounds like $Ba_2CuWO_6$ and $Sr_2MnSbO_6$ that host Jahn-Teller ions like $Cu^{2+}$ and $Mn^{3+}$ exhibit a tetragonal distortion [8]. The magnetization of the A-site doped samples is given in figure 3. It is seen that the Neel temperature of all the samples are nearly the same temperature showing that the magnetic transition temperature is insensitive to the doping in A-site. This is in agreement with the band structure calculations and UV-visible diffuse reflectance spectroscopy, measurements in the double perovskites that contain B½ ions like $Nb^{5+}$, $W^{6+}$ etc. ($nd^0$ configuration) where the substitution in the A-site does not change the band gap significantly [9].

All the three compounds do not show a linear behaviour in $1/\chi$ vs T in the measured temperature region (up to 400K). However, it is seen that the persistence in magnetic correlations above $T_N$, (seen as hump in the $1/\chi$ vs T plot) vary as we vary the ion at the A-site. The extent of this correlation is lowest for BSFNO. This is because among the three compounds, BSFNO has the highest variance in A-site cation radii [10], which could have hindered the formation of these short-range magnetic correlations.

In conclusion, we have studied the magnetic properties of $Ba_2FeNbO_6$ and $Ba_2MnNbO_6$. Both compounds are antiferromagnetic insulators with $T_N$=25K and 12K respectively. ESR measurements at room temperature show that the $Ba_2MnNbO_6$ is not Jahn-Teller distorted as expected. It is also seen that the Neel temperatures of the $A_2FeNbO_6$ (A=Ba, Sr, BaSr) compounds do not vary significantly. However variations in the average A-site ionic radius influences the formation of short range correlations that persists above $T_N$

## Acknowledgments

MSR would like to thank BMBF (Bundensministerium für Bildung und Forschung), Germany for the award of a joint collaborative Indo-German project. We thank

N. Sivaramakrishnan, RSIC, IIT, Madras and S. Angappane, Dept. of Physics, IIT, Madras for assistance with the ESR measurements.

# Figure Captions

**FIGURE 1.** XRD patterns of $Ba_2FeNbO_6$, $Sr_2FeNbO_6$, BaSrFeNbO6 and $Ba_2MnNbO_6$

**FIGURE 2 :** Temperature variation of magnetization for $Ba_2FeNbO_6$ and $Ba_2MnNbO_6$.Inset shows the derivative w.r.t. to T. Vertical arrows points to $T_N$ and horizontal arrow points to the minima in dM/dT vs T plot. The left inset in the $Ba_2MnNbO_6$ panel shows the low temperature behaviour.

**Figure 3**. Left Panel: Magnetization curves of $Ba_2FeNbO_6$, $Sr_2FeNbO_6$, $BaSrFeNbO_6$ and $Ba_2MnNbO_6$. The arrow points to $T_{N.}$ Right Panel corresponding $1/\chi$ vs T plots. The arrow indicates the magnetic correlations above $T_N$. Solid line shows the Curie-Weiss fit.

**FIGURE 4:** Magnetic hysteresis curves at 5K for $Ba_2FeNbO_6$ and $Ba_2MnNbO_6$.Inset shows the low field response showing the weak ferromagnetic like behaviour for $Ba_2FeNbO_{6.}$

**FIGURE 5.** Electron spin resonance spectra of $Ba_2FeNbO_6$ and $Ba_2MnNbO_6$ at room temperature (300K). The spectra are plotted relative to the resonance field $H_0$.

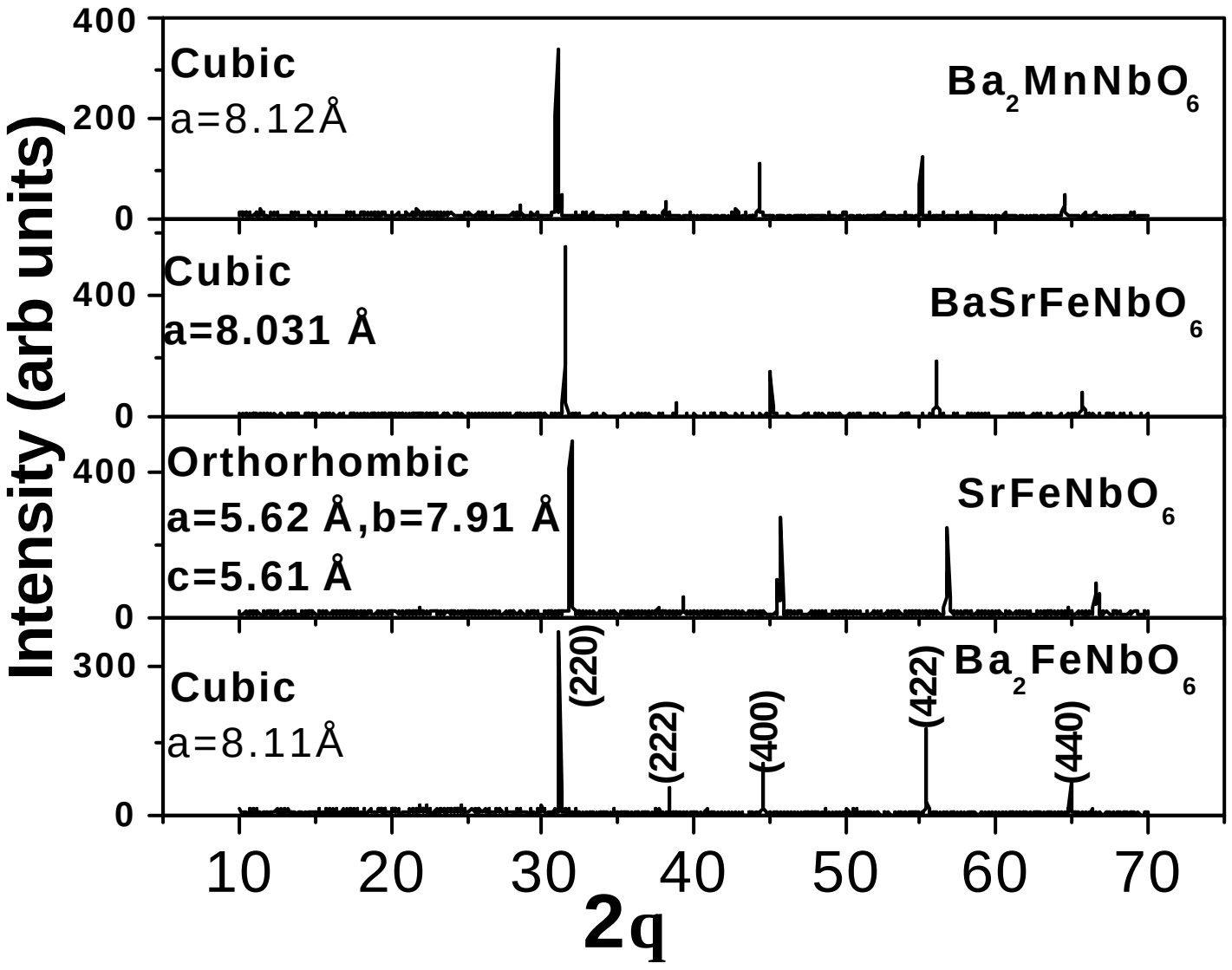

Figure 1

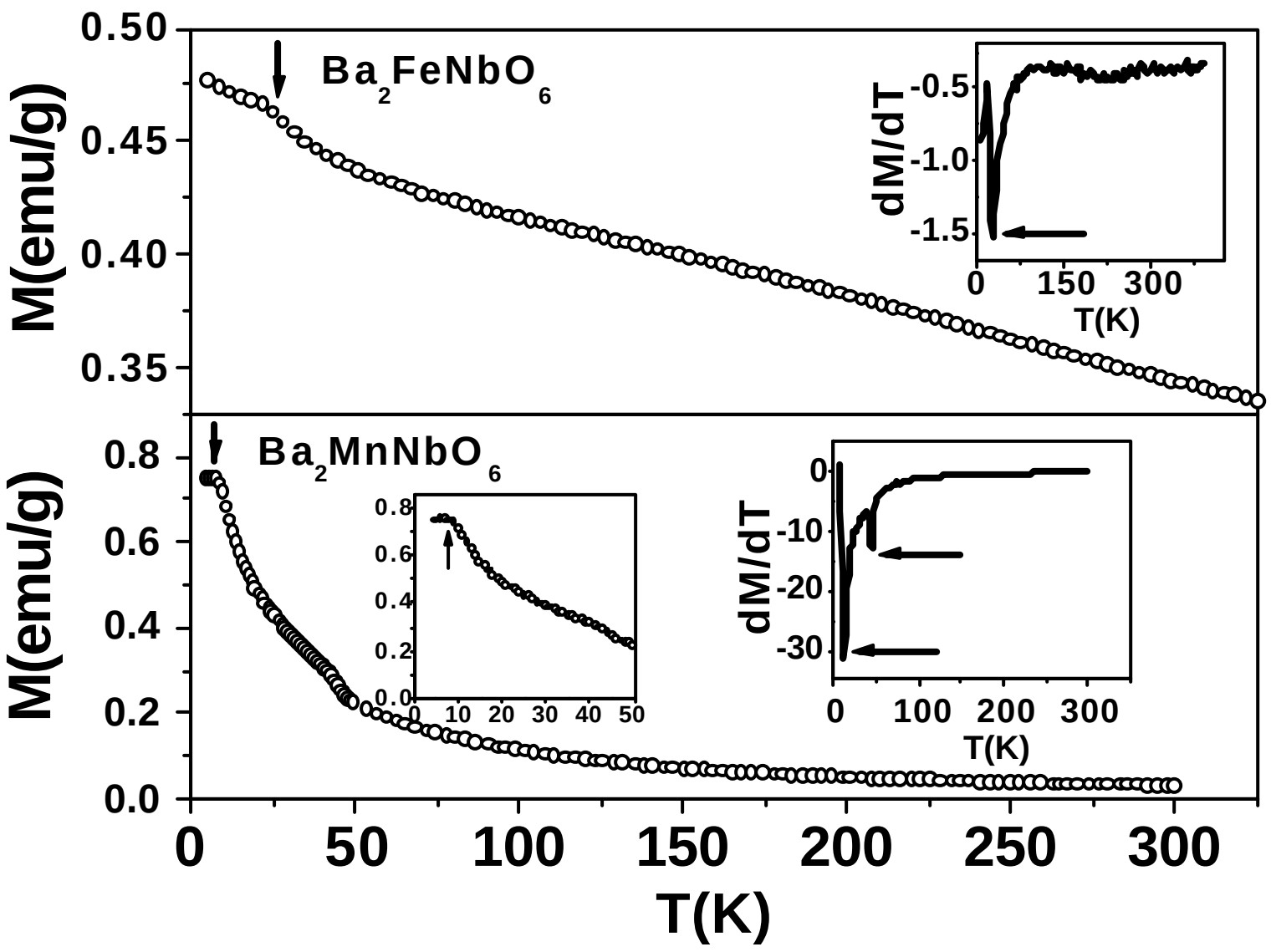

Figure2

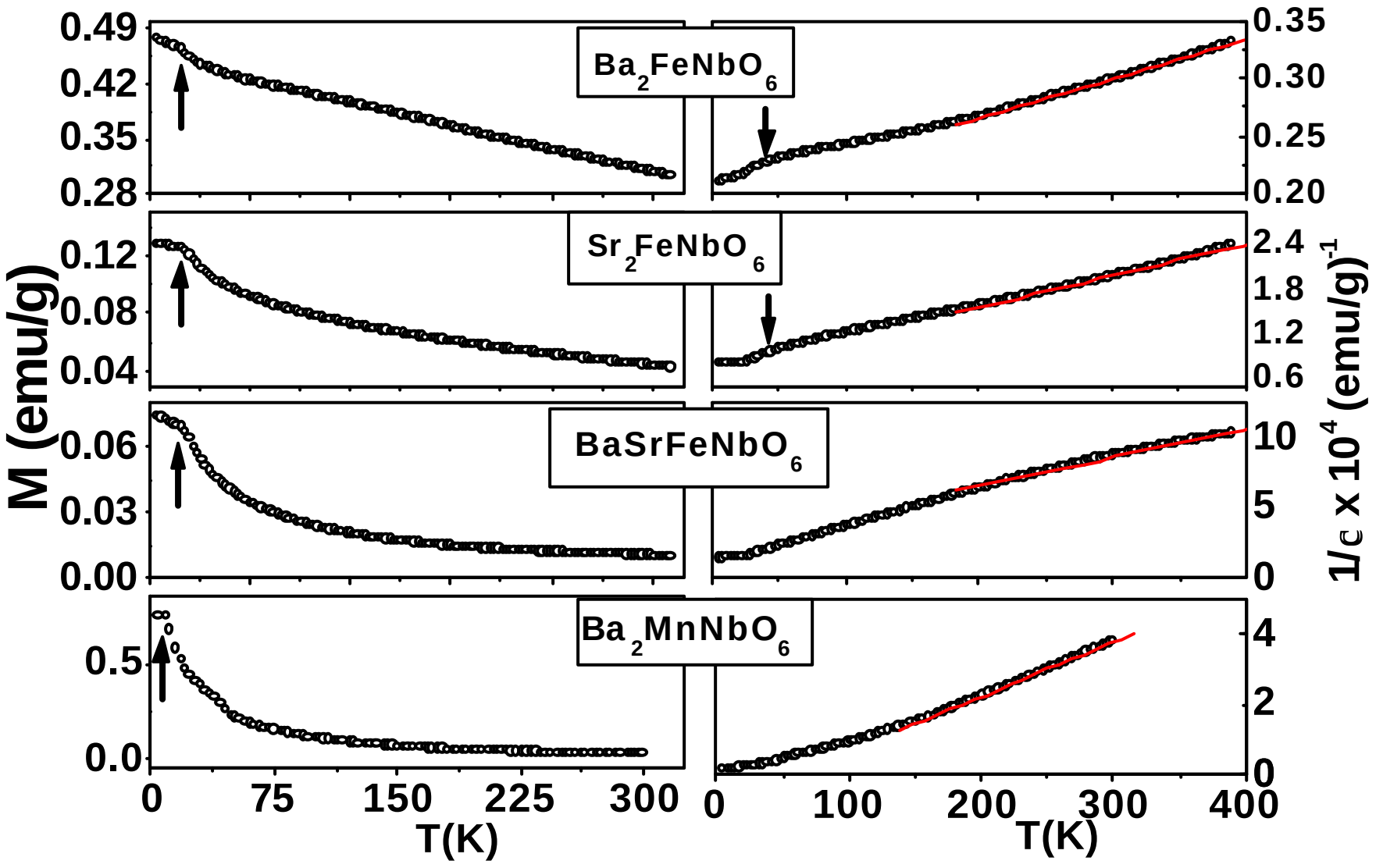

Figure3

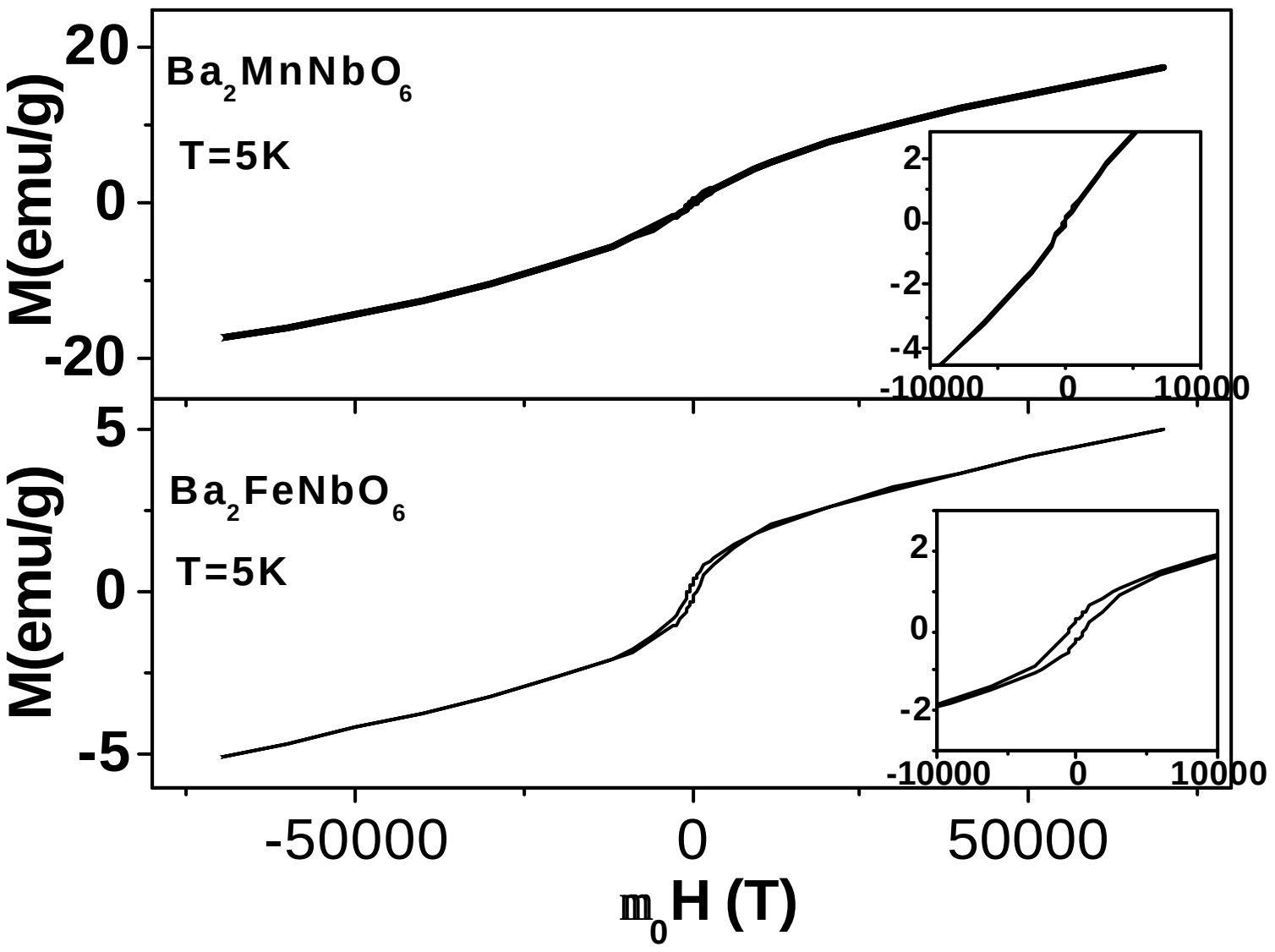

Figure4

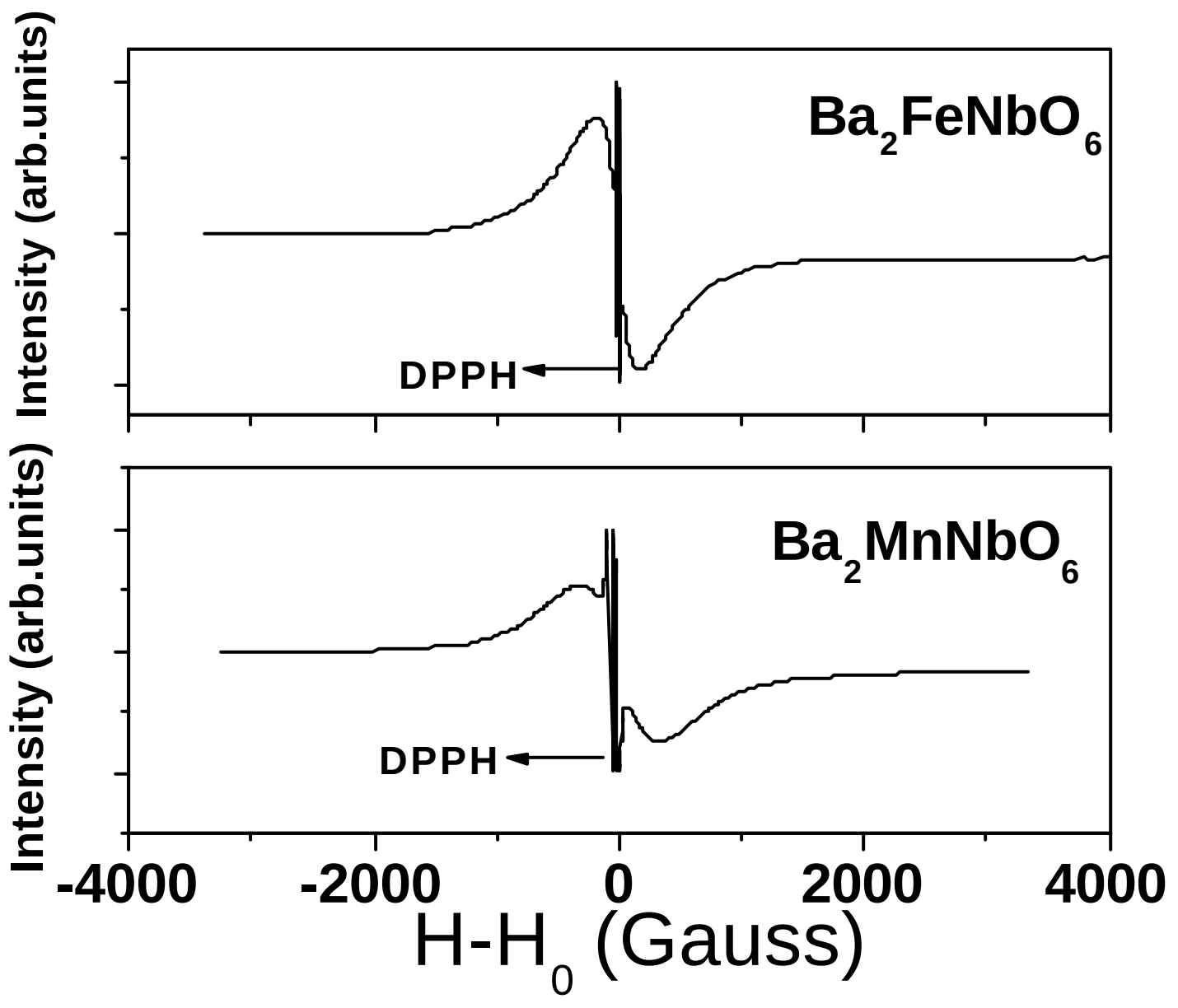

Figure5